\documentclass[aps,prl,groupedaddress,showpacs]{revtex4-1}

\usepackage{amsmath}
\usepackage{amssymb}
\usepackage{graphicx}
\usepackage{dcolumn}
\usepackage{bm}
\usepackage[usenames,dvipsnames]{xcolor}

\begin{document}
\preprint{ \color{NavyBlue} DRAFT vMAX.03 - \today}

\title{Waveguides for walking droplets}

\author{Boris Filoux$^1$\footnote{Corresponding 
author: boris.filoux@ulg.ac.be}, Maxime Hubert$^1$, Peter Schlagheck$^2$ and Nicolas Vandewalle$^1$\footnote{Website: http://www.grasp-lab.org}}

\address{$^1$GRASP, Institute of Physics B5a, University of Li\`ege, B4000 Li\`ege, Belgium.}
\address{$^2$IPNAS, Institute of Physics B15, University of Li\`ege, B4000 Li\`ege, Belgium.}

\begin{abstract}
When gently placing a droplet onto a vertically vibrated bath, a drop can bounce without coalescing. Upon increasing the forcing acceleration, the droplet is propelled by the wave it generates and becomes a walker with a well defined speed. We investigate the confinement of a walker in different rectangular cavities, used as waveguides for the Faraday waves emitted by successive droplet bounces. By studying the walker velocities, we discover that 1d confinement is optimal for narrow channels of width of $D \simeq 1.5 \lambda_F $. We also propose an analogy with waveguide models based on the observation of the Faraday instability within the channels.
\end{abstract}


\maketitle
\section{Introduction}

\begin{figure}[!h]
\centering
\includegraphics[width=7cm]{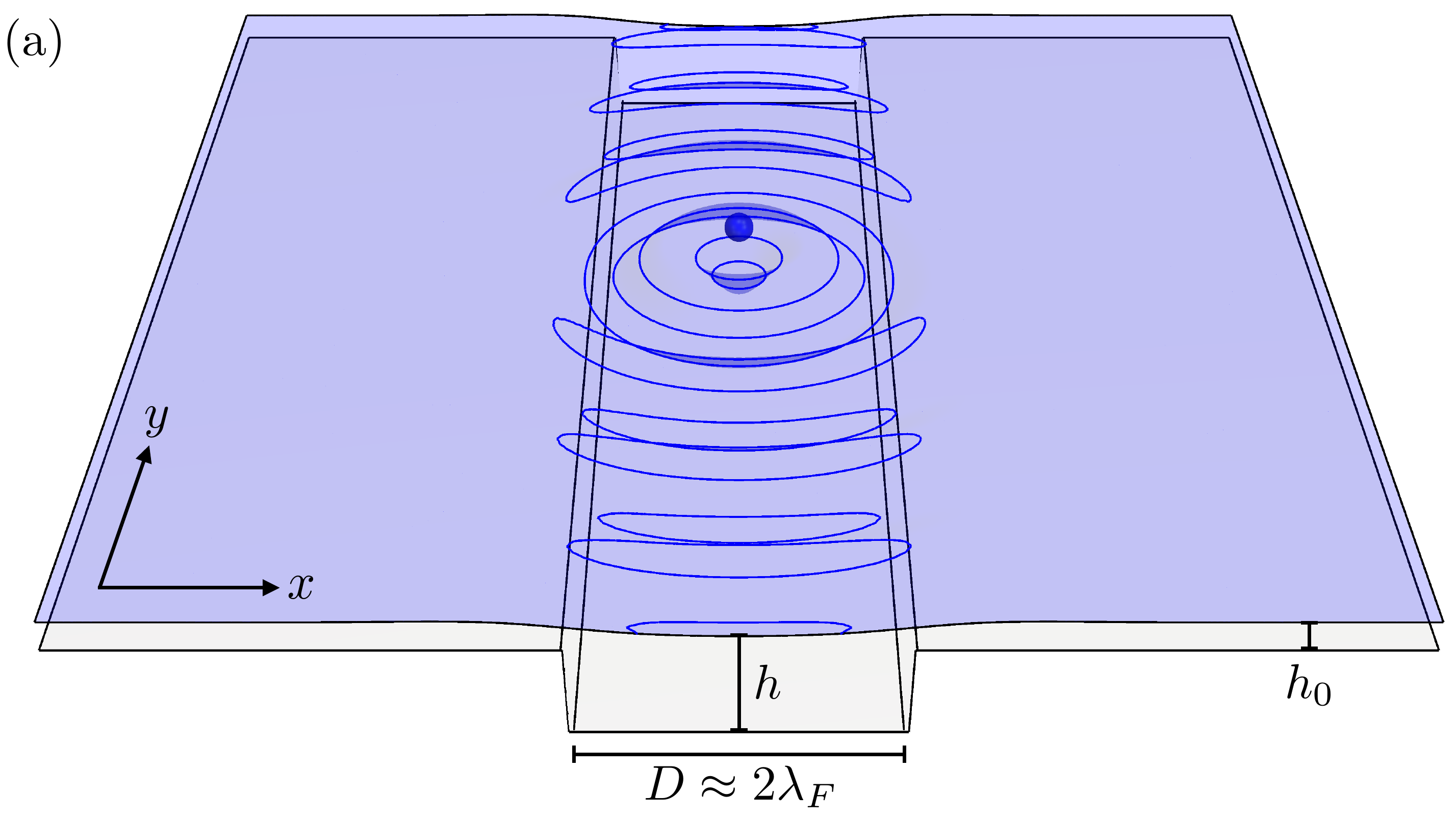} 
\includegraphics[width=9cm]{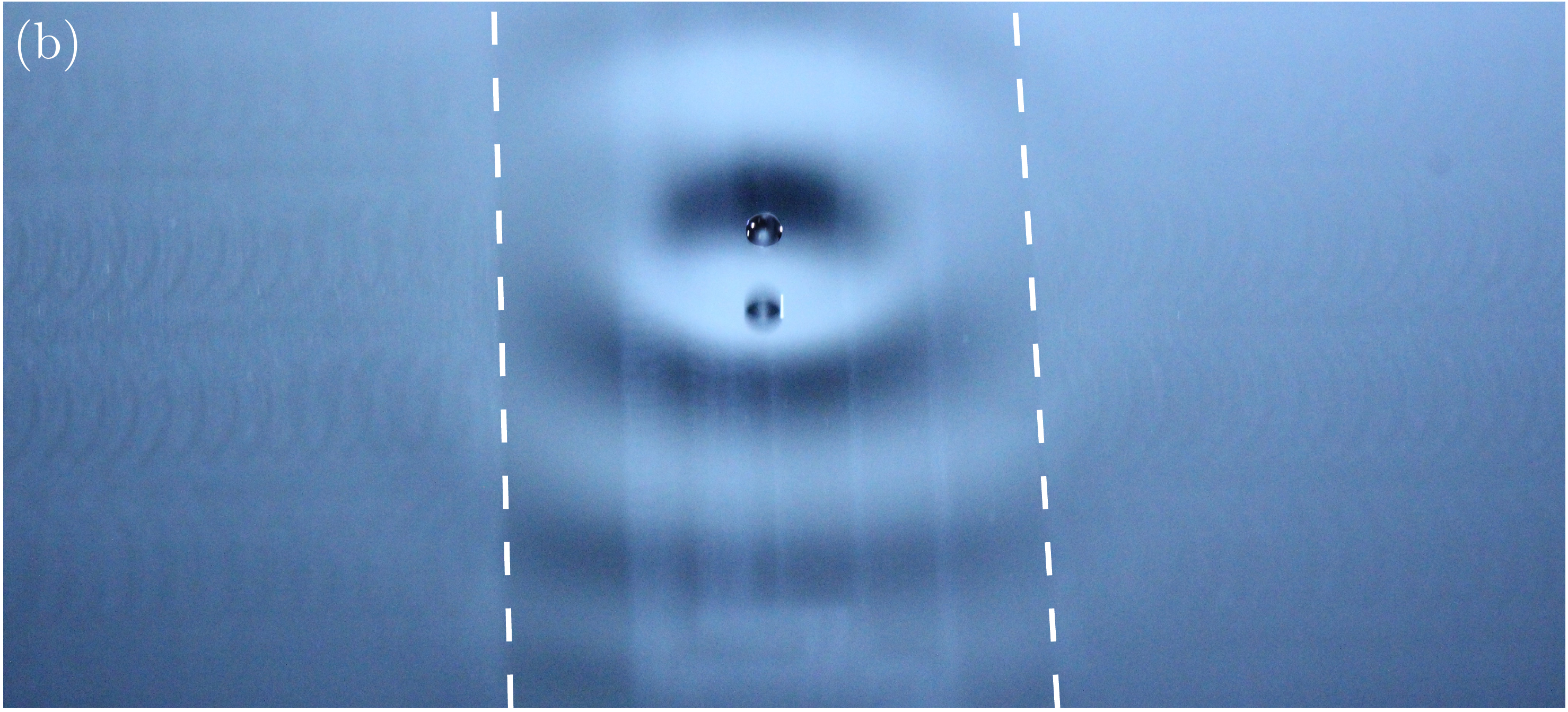} 
\caption{
(Color online) (a) Sketch of a rectangular cavity of width $D=2\lambda_F$. The liquid depth is adjusted in order to have $h = 4$ mm in the channel and $h_0 = 1$ mm elsewhere. The droplet can only walk in the channel as explained in the main text. 
(b) Picture of a walker above a channel of width $D=2 \lambda_F$. Dashed lines indicate the channel boundaries. Faraday waves excited by the droplets are seen to be strongly damped outside the cavity.
}
\label{sketch}
\end{figure}

When a droplet is gently placed at the surface of a vibrated liquid bath, inhibition of coalescence can be obtained depending on the forcing parameters \cite{Couder2005-1}. Thanks to the air layer between the drop and the bath which apply a lubrication force, the contact can be avoided. The subsequent dynamics is rich: numerous bouncing modes \cite{Wind2013}, anti-resonance \cite{Hubert2015} and rolling behaviors \cite{Dorbolo2008} have been reported. Couder and coworkers \cite{Couder2005-2} evidenced that close but below the Faraday instability threshold, the bouncing of the drop becomes subharmonic, and Faraday waves are generated at each impact. The resonant interaction between the droplet and the waves emitted leads to a self-propulsion mechanism and the droplet becomes a walker. Spectacular phenomena are therefore obtained from this peculiar wave-droplet association. For example, Perrard \emph{et al.}, using oil droplets with a ferrofluid core in a magnetic harmonic potential, obtained trajectories with quantified radii and angular momenta \cite{Perrard2014}. Using a circular coral, Harris \emph{et al.} showed that the long term statistic of the walker dynamics looks like the PDF of an electron in a corral \cite{Harris2013}. Each experimental works cited above considers 2d systems. Nevertheless, only a few investigations are based on droplets  and Faraday waves evolving in a 1d system \citep{Nachbin2015}. As circular orbits can be considered as 1d, one recent work focuses on strings of droplets propelled in annular cavities \cite{Filoux2015}. Confining droplet trajectories along 1d systems is therefore of interest from a theoretical perspective and would provide a new experimental way to manipulate such droplets.
In the present paper, we propose to study the behavior and dynamics of a single walking droplet within narrow submerged rectangular cavities, as sketched in Fig. \ref{sketch}(a) and pictured in \ref{sketch}(b). We will see that our objective to confine such droplet into 1d motions will be reached and that the walker dynamics within the channels shows interesting analogies with electromagnetic waveguides.

\section{Experimental Setup}
A container of section 180 $\times$ 180 mm$^2$ is filled with silicone oil having a kinematic viscosity $\nu=20$ cSt, a density $\rho=949$ kg/m$^{3}$ and a surface tension $\sigma=20.6$ mN/m. We use two types of containers: a square cell used for 2d comparisons and a cell with channels of 120 mm length but of different widths carved into it. We use a dispenser to generate on-demand droplets \cite{Denis2013}, with a selected diameter $2R=0.8$ mm and a precision of 1 \% . The tank is vibrated sinusoidally and vertically, using an electromagnetic shaker with a tunable amplitude $A$ and a fixed forcing frequency $f=70 \, {\rm Hz}$. An accelerometer is fixed on the vibrating plate, and delivers a tension proportional to the acceleration in order to determine the dimensionless forcing acceleration, as defined by $\Gamma = 4\pi^2Af^2/g$, where $g$ is the acceleration of the gravity. The acceleration $\Gamma$ defines the so-called ``Memory time'' \cite{Eddi2011} $\tau_M =\tau_F\Gamma_F /(\Gamma_F-\Gamma)$ which measures the time of persistence of the waves emitted by the walkers. To ensure purely vertical oscillations of the container and to avoid parasite effects on the walking dynamics of droplets, we use air cushions carriage that surround the vibrating axis, and guide its vertical motion similarly to Ref.\cite{BushShaker}.

\section{Walking Regime}
\begin{figure}[h]
	\begin{center}
	\includegraphics[width=14cm]{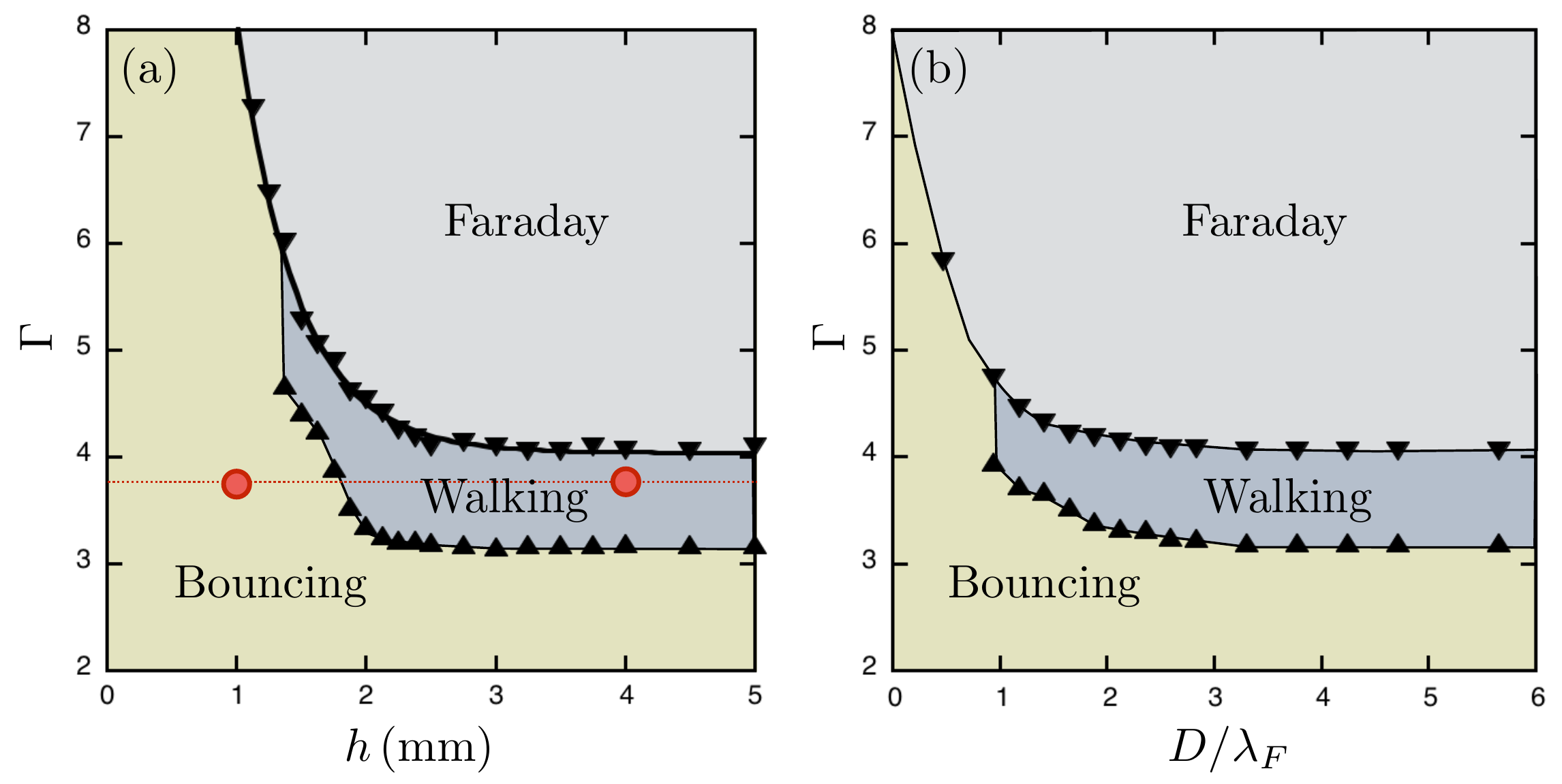} 
	\caption{(Color online) (a) For $f=70 \, {\rm Hz}$, thresholds of the Faraday instability $\Gamma_F$ (triangle down) and the walking regime $\Gamma_W$ (triangle up) as a function of the liquid height $h$. Three regimes are distinguished: Bouncing, Walking and Faraday. The horizontal dashed line corresponds to the experimental conditions ($\rm Me = 20$) and circles indicates the two depths used in this work. 
(b) Thresholds of the Faraday instability (triangle down)  and the walking regime (triangle up) as a function of the non-dimensionalized channel width $D/\lambda_F$, herein the liquid height is fixed at $h=4$ mm.}\label{gamma}
	\end{center}
\end{figure}

The droplet is bouncing above a threshold $\Gamma_B$ \cite{Couder2005-1,Hubert2015,Molacek2013b}. By increasing the acceleration above a second threshold $\Gamma_W > \Gamma_B$, the droplet becomes a walker that emits localized Faraday waves of wavelength $\lambda_F$ although the acceleration is below the Faraday threshold $\Gamma_F$ \cite{Douady1990,Bizon1998}. Walkers are only observed in the interval [$\Gamma_W$,$\Gamma_F$]. Experiments made by Eddi \emph{et al.}\cite{Eddi2009} and Carmigniani \emph{et al.}\cite{Carmigniani2014} show that modifying the depth of the fluid in the walking droplet experiment changes the value of the Faraday instability threshold $\Gamma_F$ as well as the walking threshold $\Gamma_W$ for a given frequency. In Fig. \ref{gamma}(a), we plot the evolution of both thresholds $\Gamma_F$ and $\Gamma_W$ as a function of the liquid depth $h$ for $f=70 \, \rm{Hz}$, delimiting three regimes: Bouncing, Walking and Faraday. It should be noted that this plot can be drastically modified by changing the forcing frequency $f$. Actually, we choose to work at $f=70 \, {\rm Hz}$ to obtain the largest possible interval $\Gamma_F-\Gamma_W$ for large liquid depths. This plot is quite instructive because below a critical value noted $h_c \approx 1.5$ mm, no walker is observed. However, one notices the apparition of the Faraday instability even for a small liquid depth but for large values of $\Gamma$. Finally, the cases with $h>3$ mm correspond to a deep liquid region: $\Gamma_F$ and $\Gamma_W$ are nearly independent of $h$. In our experiments, we fixed the forcing acceleration such that $\Gamma=0.95 \Gamma_F$ in the deep liquid regime, corresponding to a horizontal dashed line in Fig. \ref{gamma}(a). There, the number of impacts still emitting waves is given by the memory parameter  ${\rm Me}= \tau_M / \tau_F = 20$.

From the above results, we are able to design cavities ensuring that the droplet remains inside. Several channels were created, each one consists of a deep liquid area $h=4$ mm surrounded by a shallow liquid area $h_0=1$ mm, as illustrated in Fig. \ref{sketch}(a). These depths are measured with an uncertainty of $ \pm 0.03$ mm. For a liquid depth $h=4$ mm, we measured that the Faraday wavelength is $\lambda_F = 5.3 \pm 0.1$ mm.
Different channels have been created with a fixed length $L=120$ mm and different widths $D$. Widths from $D/\lambda_F=1/2$  up to 6 have been used. In all cases, the droplet is confined in the cavity. In Fig. \ref{sketch}(b), one can observe a walker evolving in a channel. The waves emitted by the successive bounces are seen to be strongly damped outside the rectangular cavity. Indeed, the Faraday threshold is far above the experimentally imposed forcing.
In Fig. \ref{gamma}(b), we plot the dependence of both thresholds as a function of the channel width. We note that, for large channels $ D/\lambda_F >3$, $\Gamma_F$ and $\Gamma_W$ do not vary with the channel width, and their values correspond to those obtained for a square cavity in the deep water region, as shown in Fig. \ref{gamma}(a).
In the experiments performed in smaller cavities: $D/\lambda_F < 3 $, to ensure a memory parameter of $\mathrm{Me} = 20$, one has to consequently adjust the forcing acceleration according to the evolution of the Faraday threshold within its repective channel. In our experimental protocol, $\Gamma$ is tuned so that $\mathrm{Me}$ is kept constant.

\section{Walking in cavities}
\begin{figure}[h]
	\begin{center}
		\includegraphics[width=0.48\textwidth]{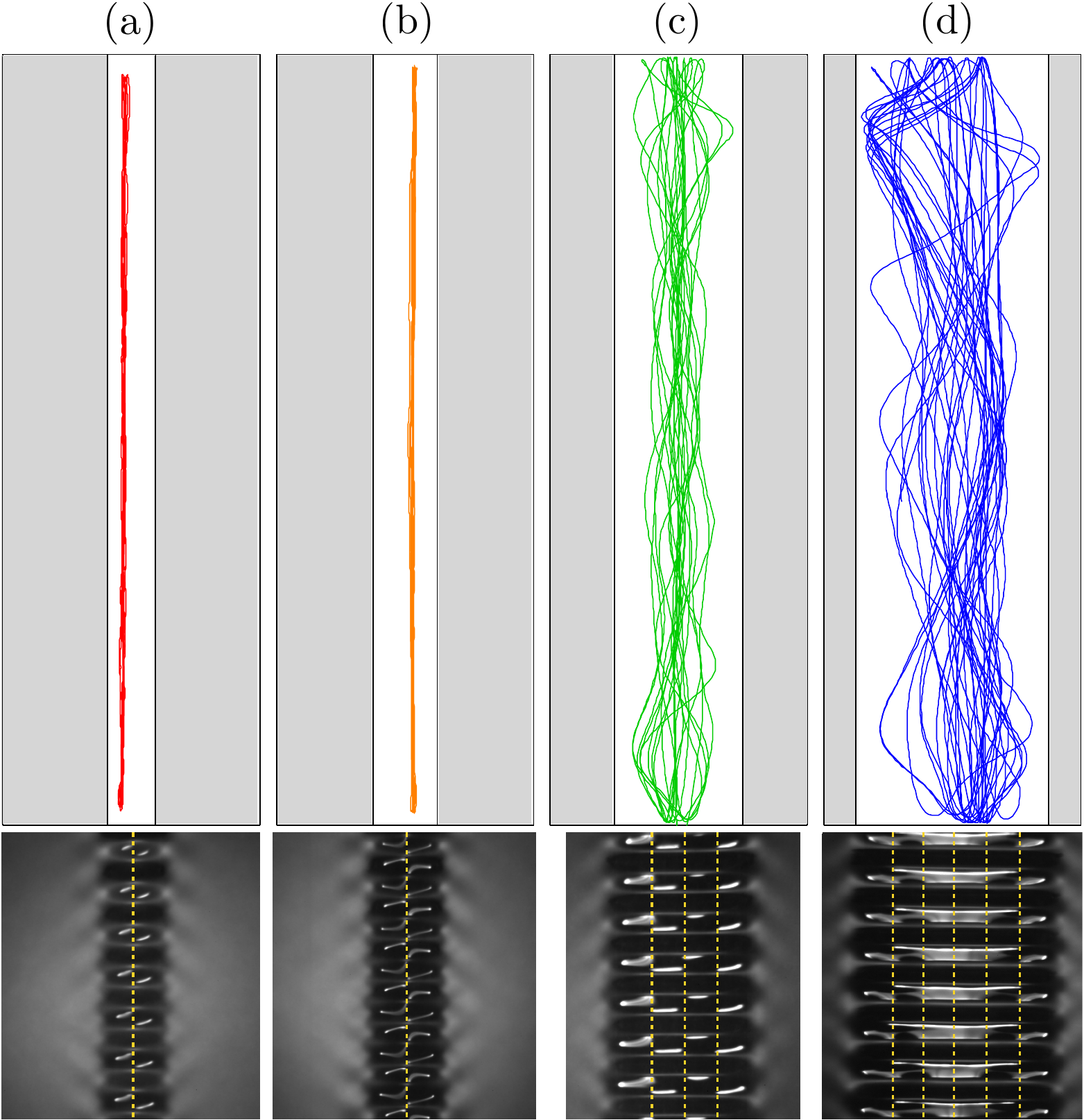}
		\caption{(Color online) (Top row) Typical trajectories of a walker in four channels: $D / \lambda_F \simeq 1.5$ (a), $D / \lambda_F \simeq 2$ (b), $D / \lambda_F \simeq 4$ (c), $D / \lambda_F \simeq 5.5$ (d).
In the first two narrow cavities (a) and (b), the walker follows a linear path. The droplet has a back and forth motion with a constant speed along the $y$-axis. In the wider cases (c) and (d), the walker is wobbling and oscillations are seen in the transverse direction, i.e. in the $x$ direction.
(Bottom row) Pictures of the Faraday pattern obtained in the corresponding upper cavities when the acceleration $\Gamma$ reaches $\Gamma_F$. One can observe evanescent waves outside the channels. The Faraday periodic pattern is only observed along the $y$ axis. The pattern corresponds to a bump along the $x$ axis, where the Faraday wavelength $\lambda_x$ slightly differs from $\lambda_y$. A secondary light source placed close to the liquid bath evidences a periodic substructure along the $x$ axis, emphasized by a yellow dashed vertical lines.} 
\label{traject}
	\end{center}
\end{figure}

Let us report various trajectories for droplets in different channels. For $D/\lambda_F<1$, the channel is able to pin the droplet which remains immobile, only bouncing on the surface. Walking droplets are only found for $D/\lambda_F\ge 1$. Typical trajectories are drawn in Fig. \ref{traject} for $D/\lambda_F \simeq \{ 1.5, 2, 4, 5.5 \}$. In narrow channels, the walker is following a rectilinear back-and-forth motion at a constant speed. When the droplet reaches an extremity, it is reflected. For wider channels, the situation is completely different since the droplet jiggles. Tortuous trajectories with several reflections on the sides of the cavity are seen. By increasing the width, the $x$ component of the droplet velocity becomes more and more important. It should be equivalent to the $y$ component for an isotropic and large system. 

\begin{figure}[b]
\begin{center}
\includegraphics[width=7cm]{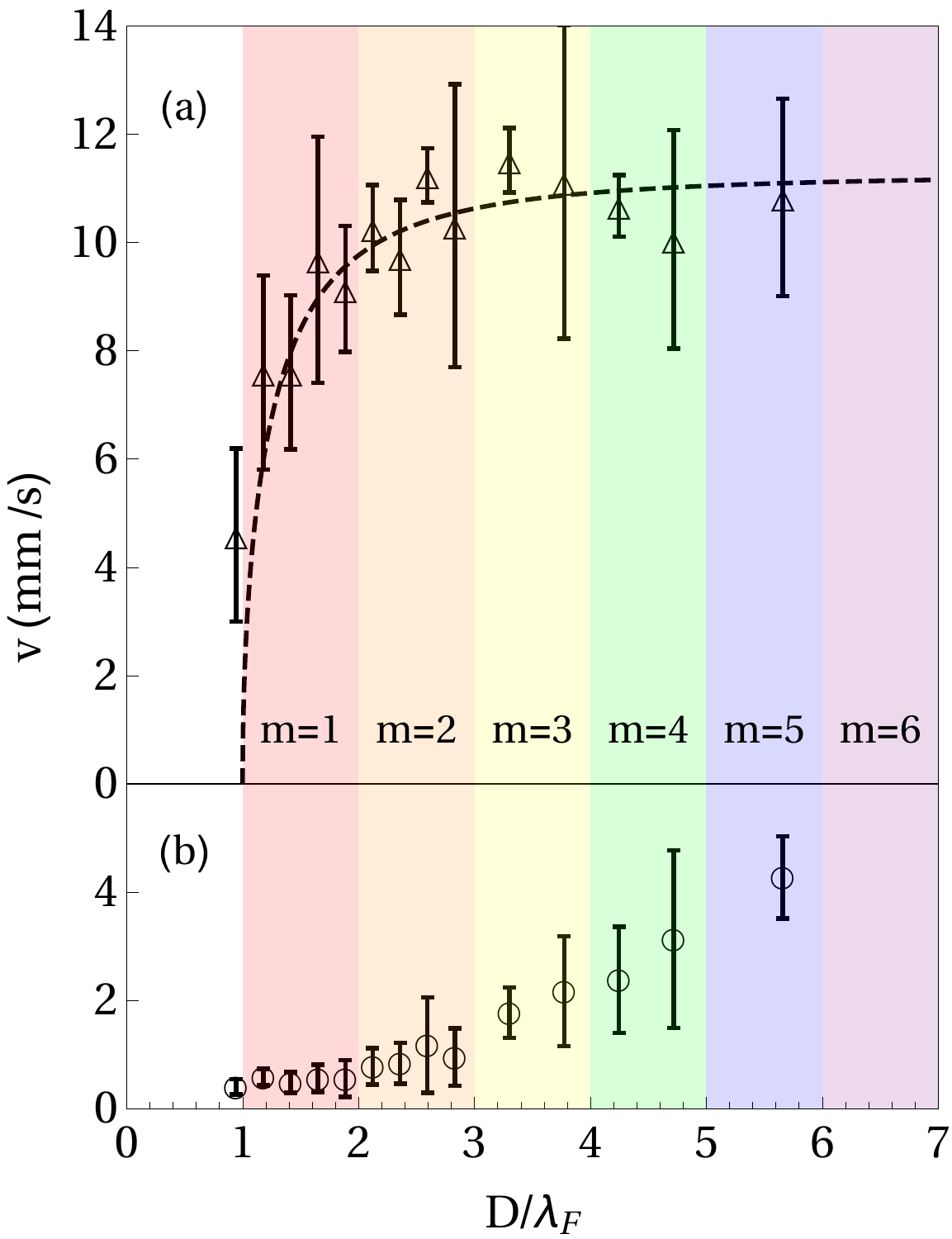} 
\caption{(a) Evolution of the average longitudinal speed $\langle \vert v_y\vert\rangle$ and (b) average transversal speed $\langle \vert v_x\vert\rangle$ of a walker as a function of the dimensionless parameter $D/\lambda_F$. The black dashed curve is a fit using Eq. (\ref{eq:WG}).   Different colors denote different propagation modes noted $m$, defined in section Discussion. Colors correspond to those of the trajectories shown in Fig. \ref{traject}.  }
\label{speeds}
\end{center}
\end{figure}

Since the successive impacts of the bouncing droplet on the surface are exciting Faraday waves, we explored the Faraday pattern into the different rectangular cavities. The forcing acceleration was cranked up until the emergence of Faraday waves. Pictures at the onset of the Faraday regime are shown in Fig. \ref{traject}. Instead of square patterns found in large tanks, we observed line patterns oriented along the transverse direction of the narrow cavities. The line spacing along the longitudinal direction corresponds roughly to $\lambda_F$.
Those patterns look similar to those obtained by Pucci and coworkers in the case of worm-like floating droplets \cite{Pucci2011,Pucci2015}. We conclude at this point that a single mode of wave propagation could exist in narrow channels since only transversal structures are observed for $\Gamma$ close to $\Gamma_F$.  Outside the channel, in the transverse direction, one can notice evanescent Faraday waves, which are strongly damped, according to our results of Fig. \ref{gamma}(a). Strictly considering line wave patterns, it is hard to understand why droplet starts to wobble for wide channels. 
A secondary light source has been placed close to the experiment in order to visualize the substructure of nodal lines which are evidenced with yellow dashed lines in Fig. \ref{traject}.
One can notice that the second light has to be tilted enough to obtain a proper visualization of the nodal lines inside the channels.
They are undulated with a wavelength close to $\lambda_F$ along the transverse direction. Please note that $\lambda_x$ is slightly different from $\lambda_y$
We note a remakable feature: the number $n_l$ of nodal lines is found to be odd in those channels. This implies that an integer (and not half-integer) number of spatial wave oscillation periods is always realized in our experiments. This peculiar property is exploited and discussed in more detail in the subsequent section.
The number of nodal lines will affect the transverse speed as shown below.
In order to quantify the walker dynamics in channels, we consider the average speed along the longitudinal and transversal directions: $\langle\vert v_y \vert \rangle$ and $\langle\vert v_x \vert \rangle$, as a function of the channel dimensionless width $D/\lambda_F$. The results are shown in Fig. \ref{speeds}(a) and \ref{speeds}(b). One observes that, along the $y$-direction, the velocity is equal to zero for channels with $D/\lambda_F\leq 1$.
As the width increases, the velocity suddenly grows to reach, for $D/\lambda\gtrsim 3$, a constant value, namely the velocity of a walker in a 2d cell without boundary effects. The behavior is different along the $x$ direction.
$\langle|v_x|\rangle$ has an almost constant value for $1 < D/\lambda_F < 2$ while it steadily increases with $D$ for $D/\lambda_F > 2$.
One expects to obtain $\langle\vert v_y \vert \rangle = \langle\vert v_y \vert \rangle$ for infinitely large channels since one recovers, in this case, the 2D walker dynamics. 
The evolution of $\langle\vert v_y \vert \rangle$ with $D$ is described later in this article thanks to a electromagnetic wave-guide analogy. The evolution of $\langle\vert v_x \vert \rangle$ seems to be linked to the Faraday instability as discussed before. Indeed, when only one nodal line is observed   (i.e. $n_l=1$, which corresponds to one wavelength in the transversal direction), $\langle\vert v_x \vert \rangle$ is constant and only changes when two new nodal lines appears (i.e. $n_l=3$, which corresponds to the emergence of a second wavelength along the transversal direction) for $D/\lambda_F = 3$.
Note that the subsequent growth of $\langle|v_x|\rangle$ is slightly less pronounced for $2 < D/\lambda_F < 3$ than for $D/\lambda_F > 3$. We attribute this to the fact that in this particular interval of channel widths a single nodal line along the center of the channel is still most often encountered, while the theorical model developed in the next section would in principle allow also for the excitation of the subsequent mode exhibiting three nodal lines. This type of restriction is expected to become less effective for wider channels with $D/\lambda_F > 3$.

\begin{figure}
	\begin{center}
	\includegraphics[width=7cm]{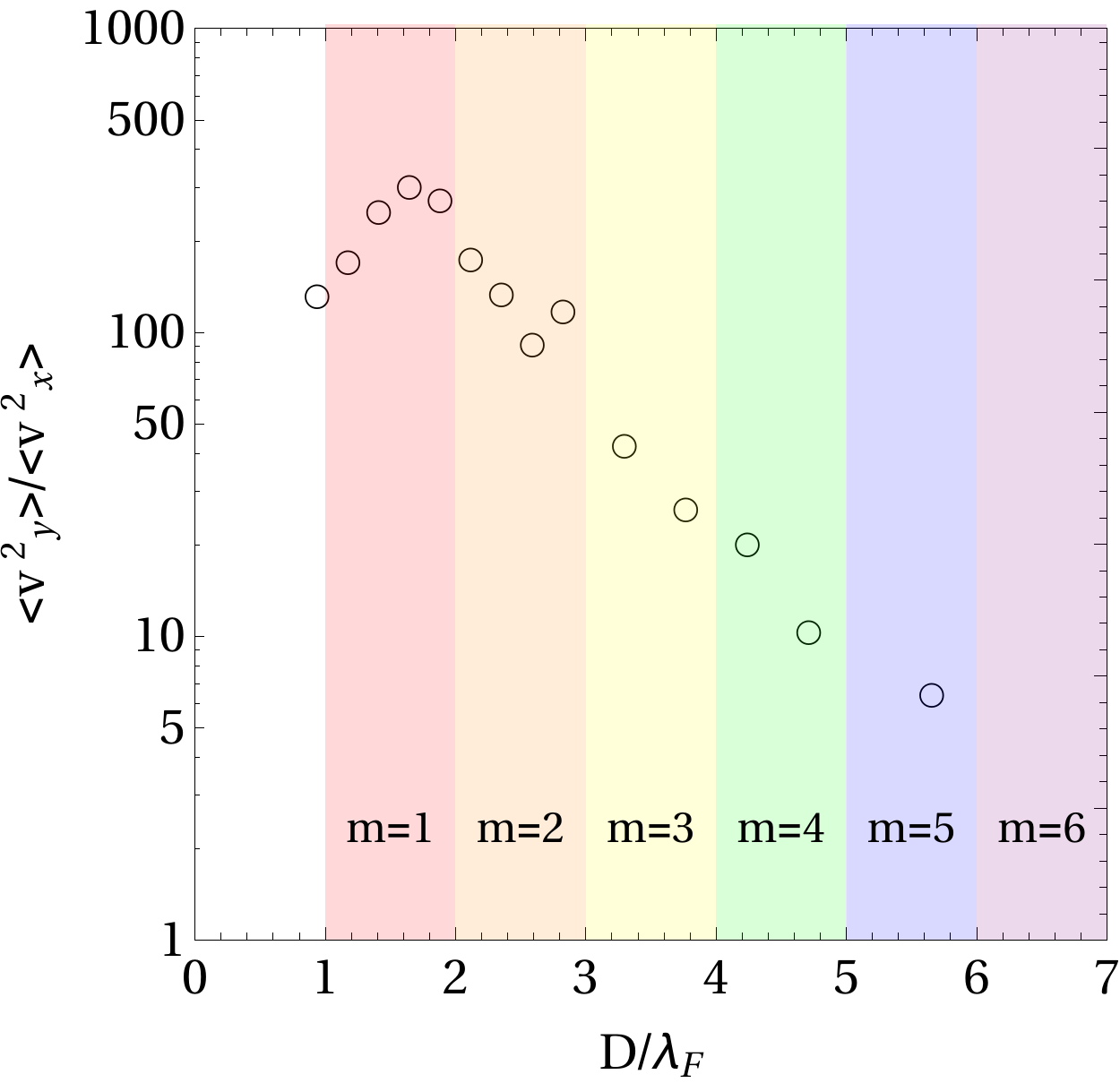} 
	\caption{(Color online) Ratio of the kinetic energies in a semi log-scale as a function of the dimensionless width $D/\lambda_F$. The ratio reaches a maximum at approximately $1.5 < D/\lambda_F < 2.5$. For very large $D/\lambda_F$ values, we expect that the ratio tends to 1 due to the equiprobability of the droplet motion in both directions.
	}\label{energy}
	\end{center}
\end{figure}

Since the main application of narrow cavities is the path control of walkers, the relevant question concerns the choice of the optimal width $D$ for limiting the speed fluctuations along the transverse axis. Figure \ref{energy} presents in a semi-log plot the ratio of the average kinetic energies measured along the $x$ and $y$ directions, i.e. the ratio $\langle v_y^2 \rangle / \langle v_x^2 \rangle$, as a function of $D/\lambda_F$. As expected, a peak is observed around  $1.5 \leq D/\lambda_F \leq 2.25$. It should be noticed that the ratio reaches values above 100.
One can consider these range as optimal widths to perform 1d experiments with walkers, in our conditions.

\section{Discussion}
The above observations can be qualitatively captured by means of an approximate theory of the Faraday waves that are emitted by the bouncing droplets. In a confinement-free configuration, a single bounce of a droplet at the position $\mathbf{r}_0 = (x_0,y_0)$ gives rise to a periodically
oscillating surface wave profile of the form \citep{Fort2010, Molacek2013b}
\begin{equation}
  \zeta(\mathbf{r},t) = \mathcal{A} J_0(k_F|\mathbf{r}-\mathbf{r}_0|)
  \cos(\omega t / 2) e^{- t/\tau_M} \label{eq:freesurf}
\end{equation}
which decays with a rate $1/\tau_M$ that is inversely proportional to $\mathrm{Me}$. We neglect here the occurrence of a spatial attenuation $\propto e^{-|\mathbf{r}-\mathbf{r}_0|/\delta}$ of the surface waves, which is encountered in droplet experiments \citep{Fort2010}.
As we argue in more detail in Ref. \citep{Dubertrand2016}, the Bessel function of the first kind $J_0$ arising in Eq.~(\ref{eq:freesurf}) can be understood as representing the imaginary part of the retarded Green 
function $G(\mathbf{r},\mathbf{r}_0,k_F)$ that is associated with the two-dimensional Helmholtz operator, that satisfies the equation
\begin{equation}
  \left(\frac{\partial^2}{\partial x^2} + \frac{\partial^2}{\partial y^2} 
  + k_F^2 + i 0 \right) G(\mathbf{r},\mathbf{r}_0,k_F) = 
  \delta(\mathbf{r}-\mathbf{r}_0) \, .\label{eq:Helmholtz}
\end{equation}
We then have
\begin{equation}
\begin{split}
  J_0(k_F|\mathbf{r}-\mathbf{r}_0|) &= 
  - 4 \mathrm{Im}[ G(\mathbf{r},\mathbf{r}_0,k_F)] \\
  &=  \frac{1}{\pi} \int d^2 k \delta(k^2 - k_F^2) 
  e^{i \mathbf{k}\cdot (\mathbf{r}-\mathbf{r}_0)}
\end{split}
\end{equation}
which, when being inserted into Eq.~(\ref{eq:freesurf}), indicates that only those waves $\propto \exp[i \mathbf{k}\cdot (\mathbf{r}-\mathbf{r}_0)]$ persist whose wave vectors $\mathbf{k}$ satisfy the Faraday wave condition
$|\mathbf{k}|=k_F =2\pi/\lambda_F$.
The Faraday wave profile expressed in Eq. (\ref{eq:freesurf}) will be modified in the presence of sub-surface channels. We suggest that this modification can be approximately accounted for by introducing additional boundary conditions to the Helmholtz equation, see Eq. (\ref{eq:Helmholtz}), that the Green function has to satisfy. Inspecting the experimentally obtained surface wave profiles that are depicted in the lower panels of Fig. 3, we infer that the Faraday waves emitted by a droplet in a channel approximately satisfy Dirichlet boundary conditions at the lateral borders of the channel, yielding the condition $G(\mathbf{r},\mathbf{r}_0,k_F)=0$ at $\mathbf{r}=(x,y)$ with $x=0$ and $x=D$ for the Green function of the Helmholtz equation. We define the horizontal coordinate system such that the channel lies 
within $0<x<D$. They furthermore appear to exhibit a nodal line at the channel's centre at $x=D/2$ as can be seen again in Fig. 3, which is presumably required to allow for a stable droplet motion within the channel. We can then expand the Green function within the lateral channel eigenmodes
\begin{equation}
  \chi_m(x) = \sqrt{\frac{2}{D}} \sin\left( \frac{2 m \pi x}{D} \right)
\end{equation}
satisfying Dirichlet boundary conditions and 
$(-\partial^2/\partial x^2) \chi_m(x) = k_m^2 \chi_m(x)$ with
$k_m = 2m\pi /D$, and exhibiting a node at $x=D/2$.
This then yields a superposition of longitudinal waves of the form
$\chi_m(x) \exp(i k_y^{(m)}|y-y_0|)$ with $k_y^{(m)} = \sqrt{k_F^2 - k_m^2}$ for
$0 < m < D / \lambda_F$, complemented by another superposition 
of evanescent waves of the form $\chi_m(x) \exp(- \kappa_y^{(m)}|y-y_0|)$ 
with $\kappa_y^{(m)} = \sqrt{k_m^2- k_F^2}$ for $m > D / \lambda_F$ 
the influence of which can be safely neglected.
As a consequence, surface waves cannot be sustainably excited within
too narrow channels whose widths $D < \lambda_F$ are below the Faraday 
wavelength.
This in turn inhibits the walking of droplets within such narrow channels,
which is in excellent agreement with our experimental findings.
It is now straightforward to infer that the walking speed $v_y$ of the 
droplet within the channel is proportional to the effective wave number 
that characterizes the longitudinal Faraday wave pattern along the channel, 
as this particular wave number determines the slope of the surface wave
profile through which the droplet is horizontally accelerated upon impact.
In the particular case of channels whose widths satisfy 
$\lambda_F < D < 2 \lambda_F$, this effective wave number is evidently 
given by $k_y^{(1)} = \sqrt{k_F^2 - k_1^2}$ as only the lowest channel mode 
$\chi_1$ can be excited in that case.
This yields the longitudinal walking speed
\begin{equation}
  v_y = \pm v_y^{(0)} \sqrt{1 - \left(\frac{\lambda_F}{D}\right)^2}
  \label{eq:WG}
\end{equation}
where $v_y^{(0)}$ is a characteristic speed scale of the droplet without
constraints, i.e. the speed of a drop in 2d.

Equation (\ref{eq:WG}) is fitted on the data of Fig. \ref{speeds}(a), yielding as sole free parameter  $v_y^{(0)} = 11.9$mm/s, which roughly corresponds to the speed for droplets in 2d.
The fit is in good agreement with our experiments, even for $D > 2 \lambda_F$. This is attributed to the fact that longitudinal speed of the droplet is predominatly governed by the largest variation of the Faraday wave along the channel, and the latter would be provided by the lowest transverse mode $\chi_1$. The transverse speed $v_x$, on the other hand, should be determined by the highest transverse mode that enters into the superposition of modes constituting the Faraday wave profile, since this particular mode provides the largest wave profile variations across the channel. We therefore should generally expect $|v_x| \propto m$ where $m = m_{max}$ is the largest integer that satisfies $m < D / \lambda_F$. This expectation is roughly satisfied as we can see in Fig. 4(b). There are, however, significant deviations due to the fact that in practice this highest transverse mode $\chi_{m_{max}}$ hardly ever participates in the experimentally observed Faraday wave profile, as we saw in Fig. 3. A more elaborate theoretical framework, taking into account the decay of Faraday waves outside the channel in a more quantitative manner, is certainly needed in order refine the model under consideration and thereby obtain a more detailed understanding of the behaviour of walking droplets in channels.
The properties of the walkers dynamics reminds the propagation of light in rectangular electromagnetic wave-guides. In particular, one may notice the ``Ray-Optics approach'' used to understand the wave propagation in a guide, where the rays of light bounces on each wall. Let us compare both systems: wave-guides are usually made of two materials with refraction index $n_1$ and $n_2$. The light propagates in the medium of index $n_1 < n_2$. Therefore, the light is focussed in the guide thanks to total reflection. In the walking droplet experiments, the two ``materials'' are the regions of different fluid depths which alter the propagation of the Faraday waves, as seen in Fig. \ref{traject}. The walker is constrained within the channel thanks to the differences between the Faraday thresholds. More interestingly, the analogy can be pushed further by comparing the velocity of the walker to the group velocity of the light propagating within an electromagnetic guide.
One can compare here the velocity of the walker to the group velocity of the light propagating within an electromagnetic guide. Indeed for TE1 wave guides, the light group velocity is given by Eq. (\ref{eq:WG}).
Nevertheless, some differences occurs. Indeed, the Faraday instability auto-adapts to the channel width where, in the electromagnetic case, the wavelength is an experimental fixed parameter. Therefore, the comparison breaks down for large channels which should be compared to TE$_n$ wave guides. Furthermore, this waveguide analogy does not give any explanation for the evolution of $\langle \vert v_x\vert \rangle$ with the width of the channel. Finally, the wave emitted by the droplet are standing waves while electromagnetic wave-guides consider only propagating waves.
Analogies should therefore be taken with caution.

\section{Summary}

In summary, we evidenced that it is possible to confine and transport walking droplets along 1d channels.
While the longitudinal motion is dominated by a single mode, a fine structure composed of $m$ modes is observed in the transverse direction. We discussed the similarities and differences with waveguided systems. We have shown, thanks to an energetic study, that the optimal width maximizing the longitudinal speed is around $2 \lambda_F$ such that applications can be designed. In particular, annular cavities have been investigated where droplets share a common coherent wave that propels the group at a velocity faster than a single drop \cite{Filoux2015}. In addition, upon shedding light on the following question: how straight the walker's motion can be in a 1d channel, this study completes the works based on walkers in confined geometries. Among them, one can note the numerical study of Gilet \cite{Gilet2014} and the experimental work of Harris \textit{et al.} \cite{Harris2013} where the authors focused on the statistics of walkers in a corral cavity and cavity modes. We plan also to study branched structures in order to manipulate walkers and organize collisions or self-interferences. Such narrow cavities are therefore the beginning of many original investigations. 

\vskip 0.2 cm
{\bf Acknowledgments} -- This work was financially supported by the Actions de Recherches Concert\'ees (ARC) of the Belgium Wallonia-Brussels Federation under Contract number 12-17/02.

\end{document}